\documentclass[preprint,11pt]{aastex}
\usepackage{natbib}
\usepackage{epsfig}
\usepackage{pdfpages}
\usepackage{footnote}

\makeatletter
\renewcommand\@biblabel[1]{\hspace{-\labelsep}}
\makeatother

\begin{document}

\title{Detection of Echoes in PSR B1508+55 at frequencies below 100 MHz using the LWA1}

\author{K. Bansal\altaffilmark{1}, {G. B. Taylor\altaffilmark{1}},
{Kevin Stovall\altaffilmark{2}}, \& {Jayce Dowell\altaffilmark{1}}
}

\altaffiltext{1}{Department of Physics and Astronomy, University of New Mexico, Albuquerque, NM 87131}
\altaffiltext{2}{Sandia National Laboratory, Albuquerque, NM}

\begin{abstract}

PSR B1508+55 is known to have a single component profile above 300 MHz. However, when we study it at frequencies below 100 MHz using the first station of the Long Wavelength Array, it shows multiple components. These include the main pulse, a precursor, a postcursor, and a trailing component. The separation of the trailing component from the main peak evolves over the course of a three year study. This evolution is likely an effect of the pulse signal getting refracted off an ionized gas cloud (acting as a lens) leading to what appears to be a trailing component in the profile as the pulsar signal traverses the interstellar medium. Using this interpretation, we identify the location and electron density of the lens affecting the pulse profile.
\end{abstract}

\section{Introduction}

Although individual pulses from pulsars are quite varied, pulsars typically have a stable integrated pulse profile over timescales of years to decades \citep{helf75,liu12}. This is one of the reasons why they are also considered to be one of the best clocks in nature and are being incorporated for the search of gravitational waves by the pulsar timing arrays (EPTA\footnote{European Pulsar Timing Array \citep{stappers06}}, PPTA\footnote{Parkes Pulsar Timing Array \citep{manchester13}} and NanoGrav\footnote{North American Nanohertz Observatory for Gravitational Waves \citep{laughlin13}}). Despite this stability, numerous pulsars have been observed to manifest a continuous pulse profile evolution. 

Pulse variations can be classified on the basis of the length of the timescale. Short scale variations, such as nulling and mode changes occur on timescales ranging from a few pulse periods to hours and days \citep{wang07}. Longer timescale variations, such as intermittent pulsars where the spin down rate undergoes a quasi-periodic cycle between phases, occur on the range of months to years. This study concerns longer timescale variations and, therefore, only those will be discussed in more detail below. Some of these variations have been attributed to changes in the pulsar magnetosphere \citep{hobbs10}. Additional examples of pulse profile variation on long timescales have been observed in binary systems due to geodetic precession or free precession. For example, \cite{cordes08} mention a specific case of an asteroid in orbit around a pulsar, which may affect the pulse profile either due to geodetic precession or if debris from an asteroid interacts with the pulsar magnetosphere. Moreover, propagation effects as the signal travels through the turbulent ionized interstellar medium \citep[ISM;][]{keith13} can also cause similar variations. Thus, studying pulse profile changes in radio pulsars can provide insights into the underlying physical mechanisms responsible for the observed changes and improve the precision of pulsar timing experiments.


To provide a few examples, PSR B1828--11, PSR J0738+4042, and, recently, B2217+47 have all been found to show such profile variations. PSR B1828--11 shows quasi-periodic profile variation which has been interpreted to be due to free precession of the pulsar \citep{stairs2000}. This interpretation, however, was later questioned by \cite{lyne10} when they found quasi-periodic profile changes in six pulsars (including PSR B1828--11), which are correlated with the spin-down rate, implying a relation to intrinsic processes of the pulsar. In the case of PSR J0738--4042, pulse variations were at first associated with a magnetospheric change \citep{karas11} and later due to an interaction between its magnetosphere and an asteroid \citep{brook14}. Recently, PSR B2217+47 has been found to show variations on the timescale of months to years in its pulse profile shape \citep{michilli}. The authors list three plausible causes which include free precession of the pulsar, variations in pulsar emission due to perturbations in the plasma filling the magnetosphere, and structures in the ISM creating a transient component in the form of echoes, strongly favoring the last cause. 


While studying scattering in pulsars below 100 MHz \citep{bansal19}, we noticed a variation in the pulse profile of PSR B1508+55 with time. As such a variation in PSR B1508+55 has not been reported previously in the literature, we decided to explore it in more detail. PSR B1508+55 has been studied using the Very Long Baseline Array and has been found to be a hyperfast pulsar with a speed of about 1000 km s$^{-1}$ \citep{chatterjee05}. It is known to have a single component pulse profile above 300 MHz \citep{naidu17}. In this paper, we study PSR B1508+55 at three frequency bands centered at 49.8, 64.5, and 79.2 MHz with the first station of Long Wavelength Array (LWA1). We analyze how the pulse profile evolves with frequency and time. The observations used in this study are described in Section 2 along with the data reduction methods employed. In Section 3, we analyze the changes in the different pulsar characteristics as a function of time. We discuss different ways for the origin of these variations and compare them with results from previous studies of PSR B2217+47 in Section 4. We conclude the study by summarizing our results in Section 5.

\section{Observations and Data reduction}

The LWA1 \citep{Taylor12} is a radio telescope array located near the Karl G. Jansky Very Large Array in central New Mexico. It consists of 256 dual-polarization dipole antennas operating in the frequency range of $10$ to $88$ MHz. The outputs of the dipoles can be formed into four fully independent dual-polarization beams. Each beam has two independent frequency tunings (chosen from the range of $10-88$ MHz), each tuning with a bandwidth of up to 19.6 MHz. The ability of the LWA1 to observe multiple frequencies simultaneously provides a powerful tool for studying the frequency dependence of pulsar profiles \citep[e.g.][]{Ellingson13}.


The LWA Pulsar Data Archive\footnote{https://lda10g.alliance.unm.edu/PulsarArchive/} \citep{stovall15} contains reduced data products for over 100 pulsars (Stovall et al., in prep) observed since 2011. The data products used for this study were produced by coherently de-dispersing and folding the raw LWA1 data using DSPSR\footnote{http://dspsr.sourceforge.net/index.shtml}. We used archival observations at three frequencies: 49.8, 64.5, and 79.2 MHz with a bandwidth of 19.6 MHz. For each frequency, we have used 39 archival observations between epoch MJD 57158 and 58361. Observations beginning from MJD 57158 through 57682 are an hour long and the remaining observations are half an hour long. The archival data consists of 4096 phase bins, 30 s sub-integrations time sections, and 512 frequency channels. We excise RFI using a PSRCHIVE median zapping algorithm. It removes data points with intensities more than six times compared to the median within a range of frequency channels. We then process this data in two different ways to obtain profile evolution and dispersion measure (DM) evolution over time.

\subsection{Average pulse profile evolution} We average all the frequency channels and reduce the number of phase bins to 512 to smooth the average profiles. These tasks are performed using the PSRCHIVE command \texttt{pam} \citep{van12}.  To further enhance the signal-to-noise ratio (S/N) of the trailing component, we average the data across epochs using a sliding window average with a width of three. We then remove the offset baseline from the average profiles and then normalize them by their maximum amplitude at all frequencies. These normalized profiles are cross-correlated with a reference profile such that profiles at all the epochs are aligned with each other. We assume that flux density of this source is constant and the component separation is independent of the absolute flux value of pulsar. 

We derive the reference profiles from higher frequency data at 79.2 (LWA1), 119, 139, 143, and 151 MHz (European Pulsar Network\footnote{http://www.epta.eu.org/epndb/  \citep{bilus16}}). These higher frequency profiles are fitted using a sum of Gaussians as explained in \citet{Kramer94}. In this case, we only fit for those components having S/N $> 5$ at our observing frequencies to avoid any confusion at the noise level. Note that the lower LWA1 frequencies of 49.8 and 64.5 MHz are not used for modeling due to the presence of the transient component and larger scattering at these frequencies. Using the width of the Gaussian components and relative amplitude of the postcursor (compared to the main component) at multiple frequencies, we derive both pulse width evolution and amplitude evolution parameters as a function of frequency. For more details about this technique refer to \cite{bansal19}. Using these fitted parameters, we obtain reference profiles of B1508+55 at 49.8 and 64.5 MHz. 

To demonstrate the evolution of component separation with time we need to estimate the relative location of the postcursor. For this reason we fit for the overlapping main component and the postcursor in the reference profiles. After removing the main component, we fit the trailing component using a Gaussian profile which yields its amplitude, width, and position. The position of the main component (its maximum) is subtracted from that of the trailing component for all the epochs to obtain the separation between the two.

\subsection{DM Evolution}

For the DM measurements of PSR B1508+55, the number of frequency channels in the archive files is reduced to 16 for every epoch at all three frequencies using \texttt{pam}. We then obtain profile templates by averaging profiles across all the epochs for each frequency using \texttt{psradd}, followed by smoothing the profile using \texttt{psrsmooth}. Profile templates are aligned in the phase before obtaining the TOAs. Using \texttt{pat} (a PSRCHIVE algorithm), we obtain the time of arrivals (TOAs) for these profiles. We combine these TOAs from all three frequencies into one file and fit for the pulsar's spin, astrometric position, and DM using the pulsar timing software \texttt{TEMPO}\footnote{http://tempo.sourceforge.net/} (Table 1). We use the fitted parameters and then only fit for the change in DM (DMX values) separately using TEMPO. These measure an offset of DM from a fiducial value for multiple epochs each having a specified time span. Here, we have used a time span of about three years. 

\begin{table}[h]
\begin{center}
Table 1. Ephemeris obtained for PSR B1508+55.\\
\vspace{0.1cm}
\begin{tabular}{lc}
\hline\hline
Parameter & Value\\
\hline
Right Ascension & $15^{h}09^{m}25.618^{s}(3)$\\
Declination & $55^{\circ}31' 32.19'' (2)$\\
Position Epoch (MJD) & 52275 \\
F0 (Hz) & 1.35193283734(6)\\
F1 (sec$^{-2}$) &  -9.19957D-15 (9) \\
DM (pc cm$^{-3}$) & 19.616\\
\hline
\hline
\end{tabular}
\end{center}
{Note: F0 - Pulsar Frequency; F1: Rate of change of F0;  All astrometric parameters are in the J2000.0 coordinate system.  These parameters were fit before obtaining a variation in DM (DMX) using pulsar timing. }
\end{table}



\section{Results}
In this paper, we explore various features such as the component separation, spectral index, and timing residuals of PSR B1508+55. Below we discuss these properties in more detail.

\begin{figure}[t!]
\begin{center}
\includegraphics[width=\textwidth, angle=0]{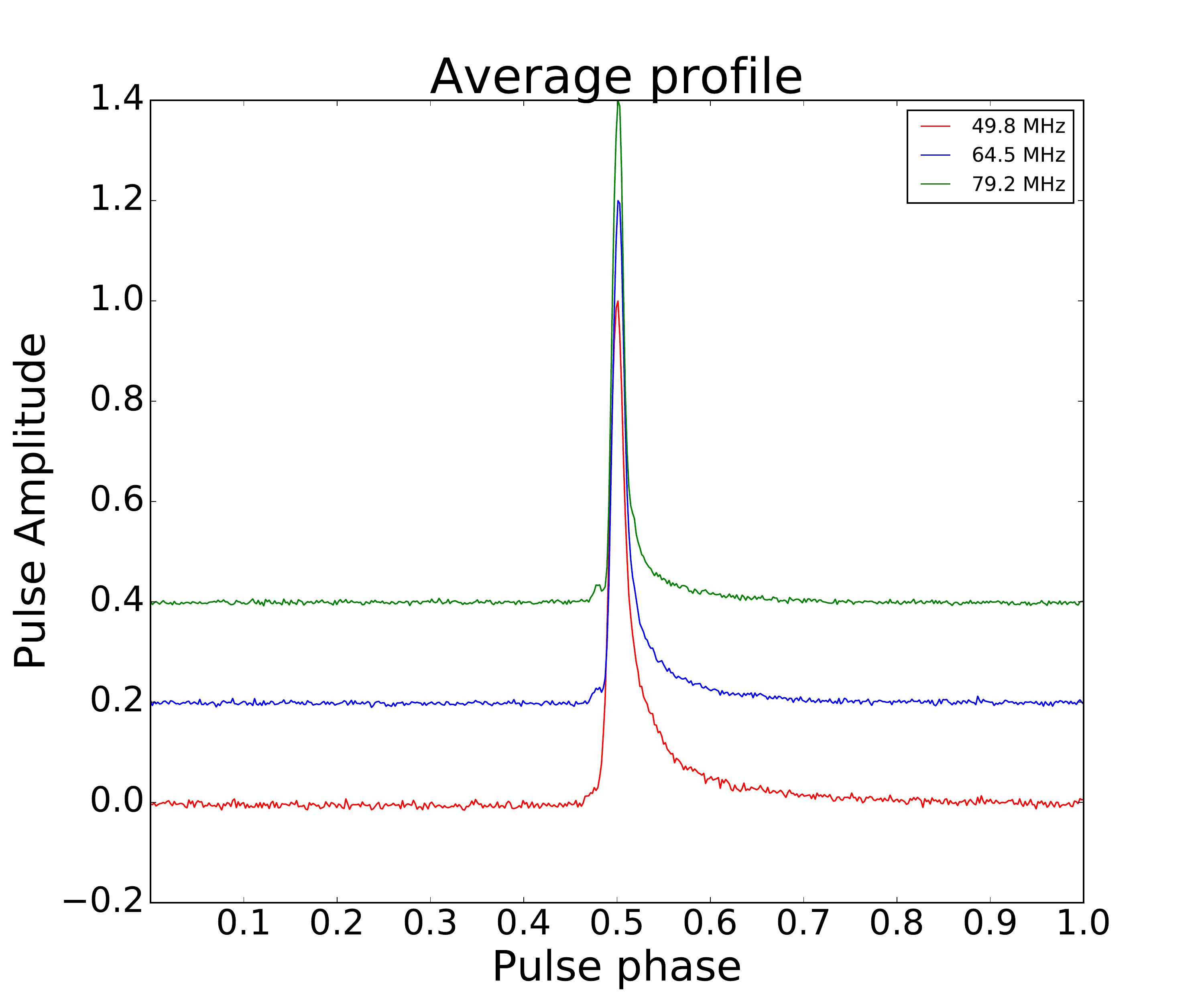}
\end{center}
\caption{Average profiles of PSR B1508+55 at 49.8, 64.5, and 79.2 MHz. All the profiles are normalized to unity and their baselines at different frequencies has been shifted in amplitude for easier visual representation.}
\end{figure}

\begin{figure}
\begin{center}
\includegraphics[width=0.82\textwidth,]{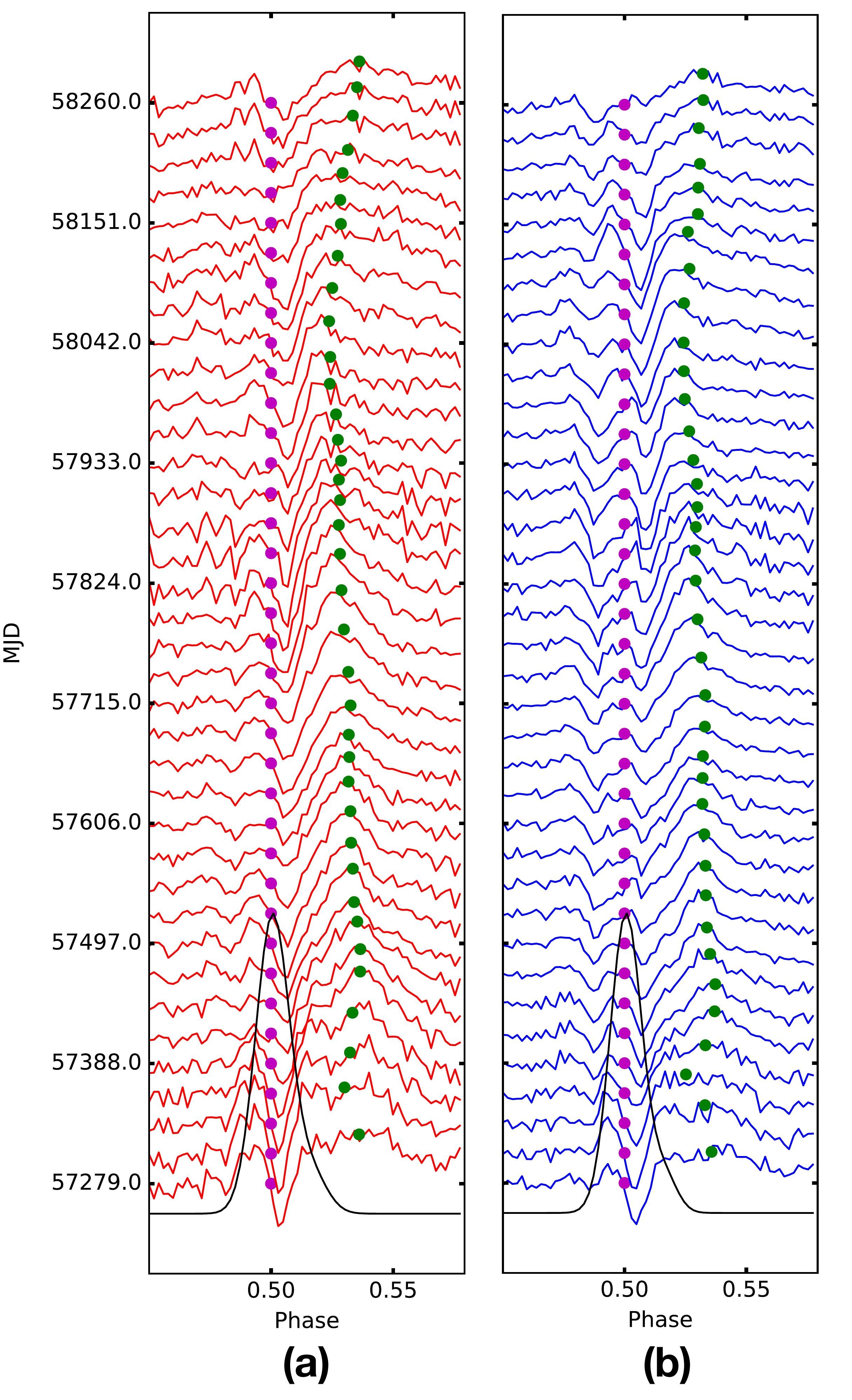}
\end{center}
\vspace{-0.3in}
\caption{Stacked residual profiles at 49.8 MHz (a) and 64.5 MHz (b) from all the available epochs. The reference profile (black) has been subtracted from the original data. The magenta points denote the location of the main component and the green dots denote the location of the trailing component, changing relative to the magenta points with time. For more details see Section 3.1.}
\end{figure}

\subsection{Profile Evolution}

Figure 1 shows averaged profiles at 49.8, 64.5, and 79.2 MHz obtained by averaging profiles across all epochs available in the LWA Pulsar archive. These profiles show a precursor component present at all frequencies with a weak indication of a postcursor component only at 79.2 MHz. On comparing this profile with those observed using LOFAR \citep{pilia16}, we note that this postcursor component is different from the trailing component. This postcursor component is not distinguishable below 64.5 MHz due to scatter broadening of main pulse. We have modeled for this postcursor and included it in the template profile (Section 2.1). The precursor component is located about 9$^\circ$ before the main component and is not visible in individual observations. The amplitude of this precursor component is about 3\% of the main component amplitude and has a S/N $< 5$, implying that this component is too weak to be modeled. Since it is not changing with time we believe it is intrinsic to the pulsar. \cite{michilli} report a similar weak precursor for PSR B2217+47 at 150 MHz. Their detection at low frequencies is likely due to a broad emission region in the pulsar magnetosphere. Both the precursor and postcursor for PSR B1508+55 have not been reported in the literature below 100 MHz.

Figure 2 (a) and (b) show stacked profiles at 49.8 and 64.5 MHz. The reference profile (black profile) has been subtracted from the original data. At individual epochs, the pulse profiles show a trailing component, which moves relative to the main component. The morphology of the trailing component is complex. It is difficult to determine the exact number of peaks in this component due to its low S/N as compared to the main component. Hence, for the sake of simplicity, we assume it to have one peak which is fit using a Gaussian. We have cut the phase after 0.58 as we do not detect any other components beyond this point. Magenta colored dots represent the position of the main component and green colored dots have been placed at the location of the fitted trailing component to demonstrate its evolution relative to the main pulse. We note that a Gaussian may not necessarily fit the maxima of the trailing components, as these are asymmetric, unlike a Gaussian profile. 

The relative amplitude of the second component for both frequencies remains mostly constant. At all the epochs, the trailing component has a lower amplitude at 64.5 MHz as compared to 49.8 MHz. We find the median value of the relative amplitude compared to the main peak to be $0.21 \pm 0.02$ at 49.8 MHz and $0.15 \pm 0.01$ at 64.5 MHz. Using these ratios, we estimate the relative spectral index of the trailing component to be $-1.30 \pm 0.18$, thus implying a steep spectrum. This is why it has poor S/N at 79.2 MHz and we were unable to include this frequency in this analysis. 

We note that this source shows temporal broadening, however, it is difficult to fit for scattering parameters over time. Its pulse profile shows a trailing component at our frequencies and its separation from the main component changes over time. At certain epochs when this trailing component overlaps with the main component, we cannot distinguish the two. We believe that this would certainly affect the estimated component separation.

Figure 3 shows the pulse separation evolution obtained from the pulse subtraction method. At both frequencies, component separation follows a similar trend which confirms the evolution of separation between the two components. Error bars on the pulse separation have been obtained from the least-square fitting algorithm. \cite{michilli} report a paper in preparation as having similar results using LOFAR. 

\begin{figure}[t!]
\begin{center}
\includegraphics[width=\textwidth, angle=0]{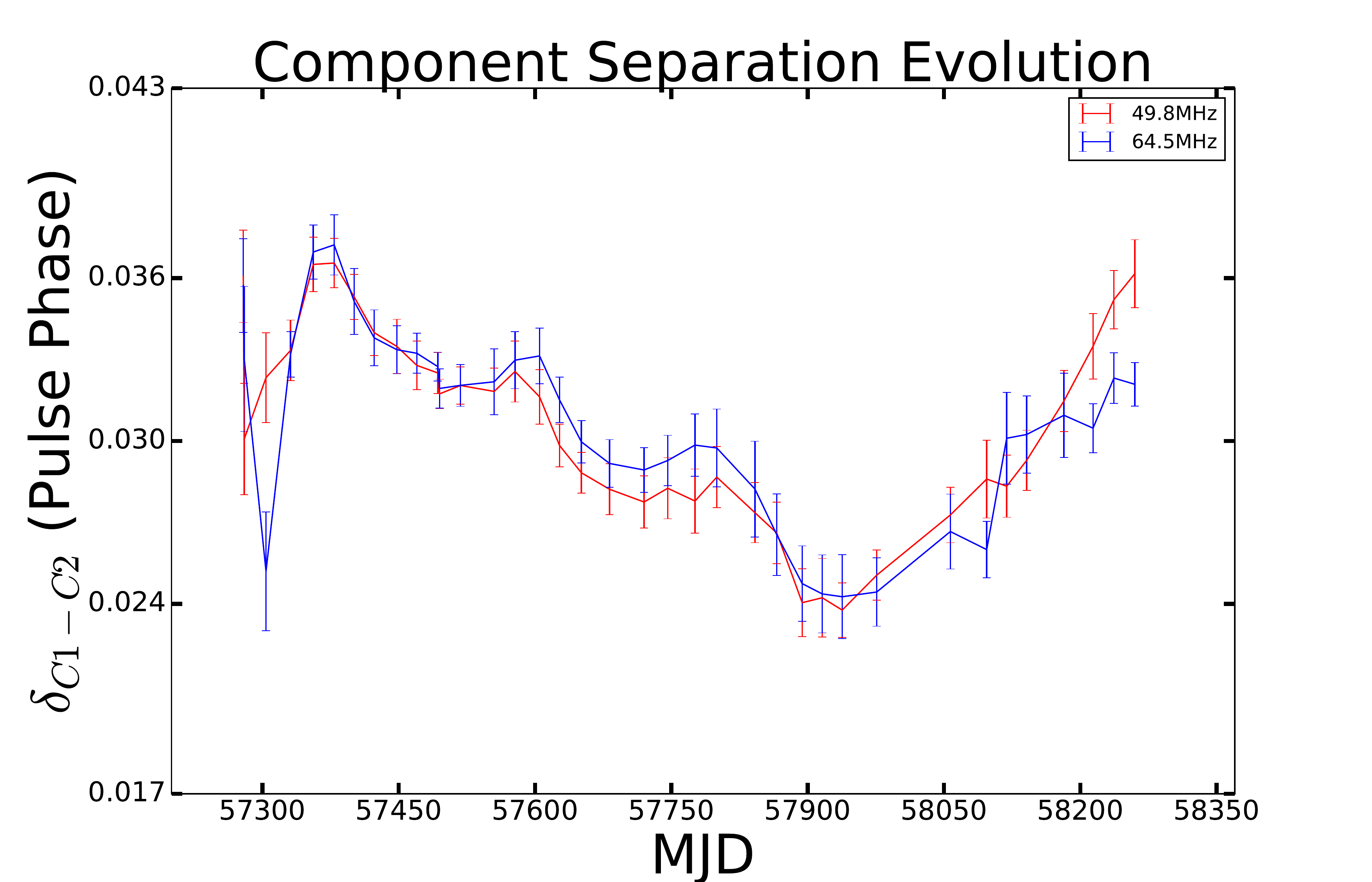}
\end{center}
 \caption{Trailing component separation obtained from subtracting the main pulse component at two frequencies: 49.8 and 64.5 MHz.} 
\end{figure}

\begin{figure}[t!]
\begin{center}
\includegraphics[trim={0 1cm 0 0},width=\textwidth, clip]{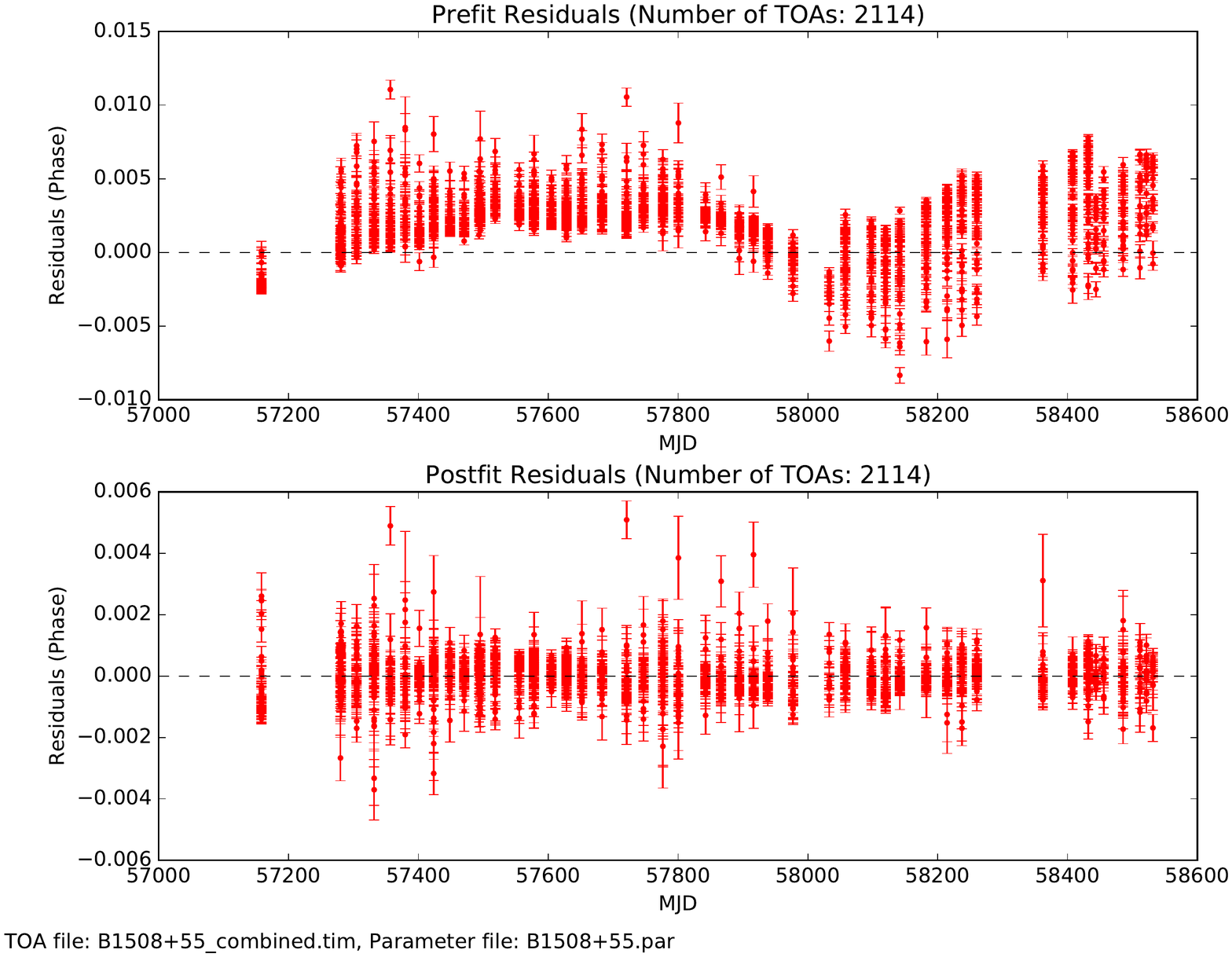}
\end{center}
\caption{Timing Residual before (top) and after (bottom) accounting for the change in DM. The y-axis changes scale by a factor of a little over two between
the pre-fit and post-fit panels.}
\end{figure}

\subsection{DM Variation}

 Our DM observations show changes in $\delta$DM across all three bands -- 49.8, 64.5, and 79.2 MHz. Figure 4 shows the timing residuals for this pulsar before and after fitting for the change in DM. Table 1 reports pulsar timing parameters. Figure 5 shows $\delta$DM values over the period of observation. The DM values start out almost constant and then there is a gradual decline with an overall change of $-6 \times 10^{-3}$ pc cm$^{-3}$ over 1000 days. We find that this variation is unlikely due to the solar wind as the minimum solar elongation angle for this pulsar is $60^{\circ}$. 
 

\begin{figure}[t!]
\begin{center}
\includegraphics[width=\textwidth, angle=0]{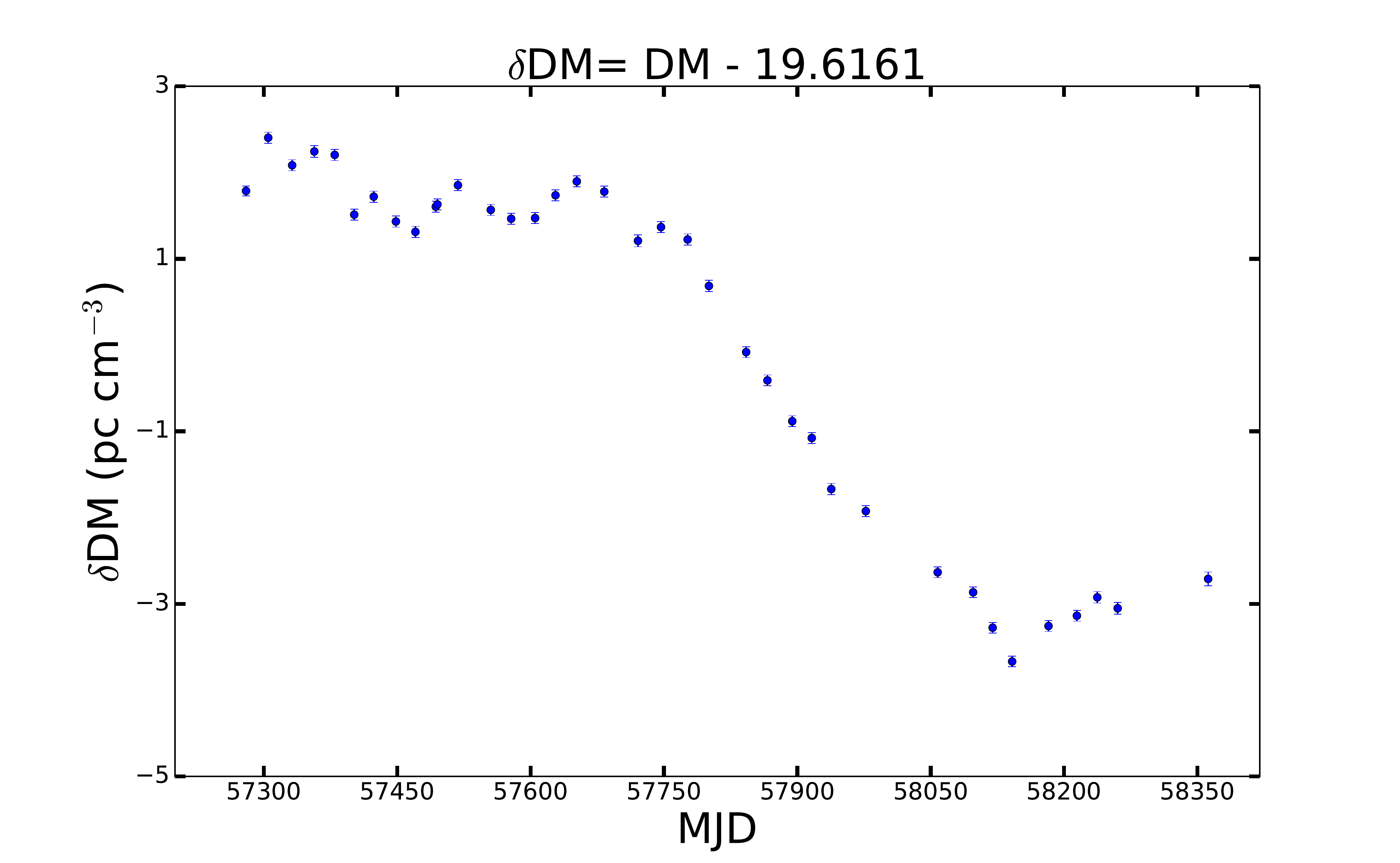}
\end{center}
\caption{Plot of $\delta$DM variation over time. $\delta$DM values represent changes in DM compared to the average value of 19.616 pc cm$^{-3}$ for PSR B1508+55. The error bars on $\delta$DM values are comparable to the symbol size.}
\end{figure}

\section{Discussion}

\cite{michilli} discussed three plausible cases for their observation of shifting components in PSR B2217+47. Now we discuss how our results compare with these cases. The first plausible reason is the free precession of a pulsar. Pulsar precession takes place when there is a misalignment between the angular momentum vector and the rotation axis. Plausible causes for this misalignment include a nonsymmetrical surface of the pulsar or a change in the magnetosphere. 

It is expected that a nonsymmetrical surface will affect the spin-down rate of a pulsar \citep{stairs2000}. We attempted to measure the spin down rate for PSR B1508+55 using a combination of three subsequent epochs. The value of the spin-down rate was constant within the errors. Additionally we measured the timing residuals (Figure 4) for PSR B1508+55 before (top) and after (bottom) accounting for the changes in DM. In the top panel of Figure 4, there is no quasi-periodic variation in the timing residuals on the timescales of our measurements as expected from a variation in spin-down rate due to precession. It shows that at each epoch the spread in the TOA residuals is due to different DM values as we have included three different frequencies. At some epochs, the TOA residuals clump together, when the DM value is equal to the DM value. We find that the TOAs residuals are due to changes in DM only. 


A second possible reason for the observed change in the pulse profile is a variation in the pulsar magnetosphere. However, PSR 1508+55 is an old pulsar ($\sim 2.3$ million years\footnote{Characteristic Age of a pulsar, $\tau_{age} = \frac{P}{2 \dot{P}}$ \citep{lyne05}.}) and, thus, we would expect the magnetosphere to be more stable as compared to a young pulsar. Hence, a change in magnetosphere are unlikely to explain our observations.


The final plausible explanation offered by \cite{michilli} includes the effect of ISM structures. When the pulsar signal travels through the ISM, structures near the line of sight (LOS) can reflect the pulse signal such that it manifests as a trailing component. Detection of the trailing component will depend on the nature of the ISM structure as it would affect both amplitude and spectral nature of the reflected pulse signal. PSR B2217+47, PSR B0531+21, and PSR B1508+55 show similar characteristics where the trailing component moves relative to the main component. However, in the case of PSR B0531+21, it is thought not to be caused by a structure in the ISM since both the pulsar and the structures causing the echo are both inside the supernova remnant \citep{lyne01}.





PSR B1508+55, similar to PSR B2217+47, shows a change in DM of $\sim 6 \times 10^{-3}$ pc cm$^{-3}$ over a period of three years (Figure 5). In the case of PSR B2217+47, \cite{michilli} associate it with the pulse profile evolution. 
Moreover, in PSR B2217+47, the minimum phase delay between the main component and the transient postcursor is 1 ms at 150 MHz while in the case of PSR B1508+55 it is 17 ms at 49.8 MHz. The difference between minimum phase delays could be due to their different location in the sky or simply the difference in observing frequencies. Since the delay due to dispersion scales as $\nu^{-2}$, we expect a smaller delay at higher frequencies.  Considering this we expect the delay at 49.8 MHz to be roughly nine times that at 150 MHz. Since the minimum delay happens at the closest approach between the pulsar and the ISM structure, the remaining phase delay of 8-ms at LWA frequencies could be due to a larger angular separation between them. Changes in DM and pulse shape due to the ISM structure should be correlated when the LOSs aligns. However, the correlation between these two parameters is not obvious here due to a time offset between the decay in $\delta$ DM and component separation of PSR B1508+55 (Figure 3 and 5 ). It appears that there may be an offset for PSR B2217+47 as well (Figures 4 and 5 in \citealp{michilli}). Similar offsets in both pulsars suggest that the moving pulse component and the DM change could be related. For example, if there are multiple structures or if a single structure has a complex density profile, it could affect both the DM and the pulse profile. However, there are several cases where pulsars show DM change over long timescales \citep{bansal19} without affecting the pulse profile and hence, these two occurring at the same time could be a coincidence. 


The scattering from the ISM structure affects the amplitude of the pulse signal, and consequently, the amplitude of the trailing component. We compare the ratio of the trailing component amplitude with that of the main component for PSR B1508+55 with PSR B2217+47. We find the value of the spectral index of the trailing component relative to the main peak to be $-1.30 \pm 0.18$. In the case of PSR B2217+47, the relative spectral index of the trailing component has been found to be $-1.60 \pm 0.03$. Both observations suggest a steep spectral nature of the trailing component.  This is probably why no observations above 300 MHz previously detected the trailing component.

The above discussion suggests that such observations are not intrinsic to the pulsar but due to the interaction of the pulsar signal with ISM. The steep spectral nature of the postcursor for both pulsars explains our observations at low frequencies. Additionally, low frequencies are highly sensitive to ISM effects which makes them more suitable for this type of study. However, it requires high sensitivity to detect such an effect. In our scattering sample of eight pulsars over a span of three years only PSR B1508+55 shows variation in its component separation over time. The reason is likely due to high scattering at these frequencies, which obscures the trailing component and makes these phenomena difficult to observe. 

\subsection{Properties of ISM structure}

Given the similarity of the profile evolution to others previously attributed to propagation effects \citep{backer01,lyne01, michilli} and the lack of evidence favoring a different interpretation, we favor a model based on a lens in the ISM to explain our observations. Considering this interpretation, we calculate the physical characteristics of this structure below.

We use the pulse delay ($\tau_{d}$) to estimate the distance of this structure from the Earth. We use the ISM scattering model proposed by \cite{michilli} and use their Equation A7. We fit for this equation using our measured component delay between MJD 57379 and MJD 57976. We use these two epochs as there are turnovers before and after them. The estimated distance between the lens and pulsar is between 186 and 316 pc (X2 in Figure 6). This makes the average distance from Earth to the lens in the ISM about 10\% in comparison to the pulsar \citep[2.1 $\pm$ 0.13 kpc; see][]{chatterjee09}, implying that the lens is relatively close to the Earth. We note that the large error in pulsar distance makes it difficult to estimate X2 more precisely.

We note that the component delay is roughly symmetric here as compared to the profile evolution presented by \cite{michilli}. If the lens causing these echoes was stationary and the pulsar was moving towards it, we would expect to see a decrement in the phase delay and then increment. However, in this case we see that first there is a slow increment in the phase delay followed by decrement and then an increment. This suggests that the lens is likely a turbulent region in the ISM. To determine the true nature of this behavior, we would need to study this source over a longer period. We also searched for nearby sources along the LOS and were unable to find any at the estimated distance from Earth.




The ISM lens consists of plasma, which has refractive properties. An average electron density ($n_{e}$ in cm$^{-3}$) of the lens can be obtained using the following relation \citep{hill05},

\begin{equation}
n_{e} = 5.4 \theta_{r} / \lambda^{2},
\end{equation}

where $\theta_{r}$ is the refracted angle in mas and $\lambda$ is the observing wavelength in m. By using the geometrical relations, we estimate the deflection angle ($\delta = \theta_{r}$) by obtaining all three sides of the triangle (Figure 6). We calculate the third side (X1) of the triangle using the extra path length measurement, traveled by light, using $X1 + X2 - D = \tau_{d} \times c$. We estimate the deflection angle to be $\sim 320$ mas. Using both observing frequencies (49.8 and 64.5 MHz), we obtain the electron density to be $\sim 40 - 100$ cm$^{-3}$. This value is of the same order of magnitude as reported in \cite{michilli} and consistent with the extreme scattering event electron density obtained in \cite{hill05}.


 \begin{figure}[t!]
\begin{center}
\includegraphics[width=\textwidth, angle=0]{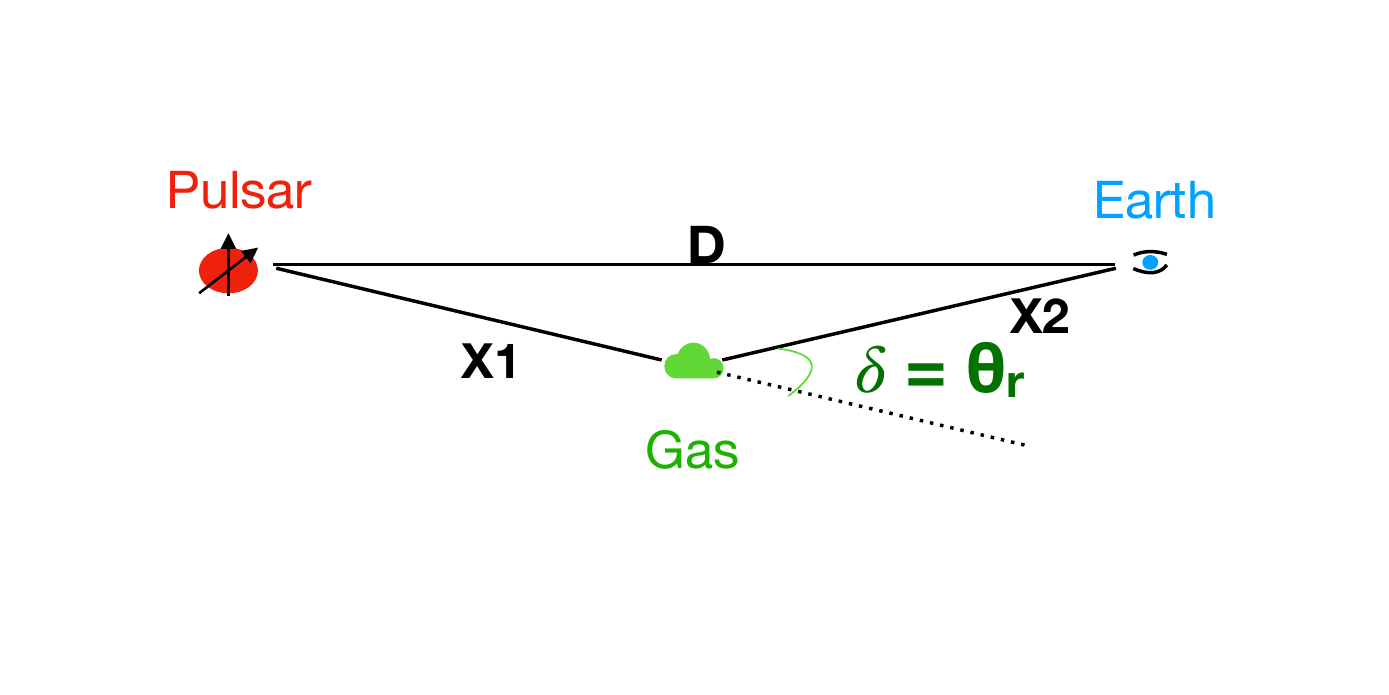}
\end{center}
\vspace{-1.0 in}
\caption{A simplified cartoon showing this scattering of pulsar signal from the ISM ionized gas cloud. The main component travels along D and the trailing component along X1-X2. The angle $\delta$ is the deflection angle, which is equal to the refractive angle. This figure has been adapted from \cite{michilli}.}
\end{figure}

\section{Conclusion}

In this paper we study the pulse profile variation in PSR B1508+55. This variation was discovered serendipitously while studying the scattering for a sample of eight pulsars using the LWA1. We compare our results from \cite{michilli}, where they have observed a similar phenomena for PSR B2217+47. Our analysis of component separation and DM evolution suggests that this observation is the result of echoes due to the presence of structures in the ISM. We estimate the distance of this structure from the Earth to be $\sim 251$ pc with an ionized density of $40-100$ cm$^{-3}$. This study suggests that observing pulsars at lower frequencies enables us to explore such structures in the ISM which could be difficult to detect at higher frequencies. We also report on the discovery of a faint precursor and postcursor component in the profile of PSR B1508+55 below 100 MHz, suggesting a possible broad cone of emission at lower frequencies.

\end{document}